\documentclass[aps,prl,showpacs,twocolumn,superscriptaddress]{revtex4}

\usepackage{amssymb}
\usepackage{graphicx}
\usepackage{dcolumn}
\usepackage{bm}

\begin{document}

\title{Fast optical preparation, control, and read-out of single quantum dot spin}

\author{A.~J.~Ramsay}
\email{a.j.ramsay@shef.ac.uk} \affiliation{Department of Physics
and Astronomy, University of Sheffield, Sheffield, S3 7RH, United
Kingdom}

\author{S.~J.~Boyle}
\affiliation{Department of Physics and Astronomy, University of
Sheffield, Sheffield, S3 7RH, United Kingdom}

\author{R.~S.~Kolodka}
\affiliation{Department of Physics and Astronomy, University of
Sheffield, Sheffield, S3 7RH, United Kingdom}

\author{J.~B.~B. Oliveira}
\affiliation{Department of Physics and Astronomy, University of
Sheffield, Sheffield, S3 7RH, United Kingdom}
\affiliation{Departamento de F$\mathrm{\acute{i}}$sica,
Universidade Estadual Paulista - UNESP, Bauru-SP, 17.033-360,
Brazil}

\author{J.~Skiba-Szymanska}
\affiliation{Department of Physics and Astronomy, University of
Sheffield, Sheffield, S3 7RH, United Kingdom}

\author{H.~Y.~Liu}
\affiliation{Department of Electronic and Electrical Engineering,
University of Sheffield, Sheffield, S1 3JD, United Kingdom}

\author{M.~Hopkinson}
\affiliation{Department of Electronic and Electrical Engineering,
University of Sheffield, Sheffield, S1 3JD, United Kingdom}

\author{A.~M.~Fox}
\affiliation{Department of Physics and Astronomy, University of
Sheffield, Sheffield, S3 7RH, United Kingdom}

\author{M.~S.~Skolnick}
\affiliation{Department of Physics and Astronomy, University of
Sheffield, Sheffield, S3 7RH, United Kingdom}

\date{\today}

\begin{abstract}

We propose and demonstrate the sequential initialization, optical
control, and read-out of a single spin trapped in a semiconductor
quantum dot. Hole spin preparation is achieved through ionization
of a resonantly excited electron-hole pair. Optical control  is
observed as a coherent Rabi rotation between the hole and charged
exciton states, which is conditional on the initial hole spin
state. The spin-selective creation of the charged exciton provides
a photocurrent read-out of the hole spin state.
\end{abstract}
\maketitle

The ability to sequentially initialize, control and read-out a
single spin is an essential requirement of any spin based quantum
information protocol \cite{Loss_pra}. This has not yet been
achieved for promising schemes based on the optical control of
semiconductor quantum dots \cite{Imamoglu_prl}. These schemes seek
to combine the picosecond optical gate speeds of excitons
\cite{Zrenner_nat,Stufler_prb,Li_sci,Patton_prl}, with the
potential for millisecond coherence times of quantum dot spins
\cite{Kroutvar_nat,Heiss_cmat,Greilich_sci}, by optically
manipulating the spin via the charged exciton. This results in a
system where the potential number of operations before coherence
loss could be extremely high, in the range $10^{4-9}$, and in a
system compatible with advanced semiconductor device technologies.
A number of important milestones have recently been reached, but
these focus on the continuous initialization of an electron
\cite{Atature_sci,Xu_prl} or hole spin \cite{Gerardot_nat},
detection
 of a single  quantum dot spin  \cite{Berezovsky_sci,Atature_natphys},
 or optical control of ensembles of $10^{6-7}$
spins \cite{Greilich_prl,Wu_prl}.


In this letter, we demonstrate sequential  triggered on-demand
preparation, optical manipulation, and picosecond time-resolved
detection of a single hole spin confined to a quantum dot, thus
demonstrating an experimental framework for the fast optical
manipulation of single spins. This is achieved using a single
self-assembled InGaAs quantum dot embedded in a photodiode
structure. The hole spin is prepared by ionizing an electron-hole
pair created by resonant excitation. A second laser pulse then
drives a coherent Rabi oscillation between the hole and positive
trion states, which due to Pauli blocking is conditional on the
initial hole spin state, key requirements  for the optical control
of a spin via the trion transition. Due to Pauli blockade,
creation of the charged exciton provides a photocurrent read-out
of the hole spin state.


\begin{figure}
\begin{center}
\includegraphics[scale=0.55]{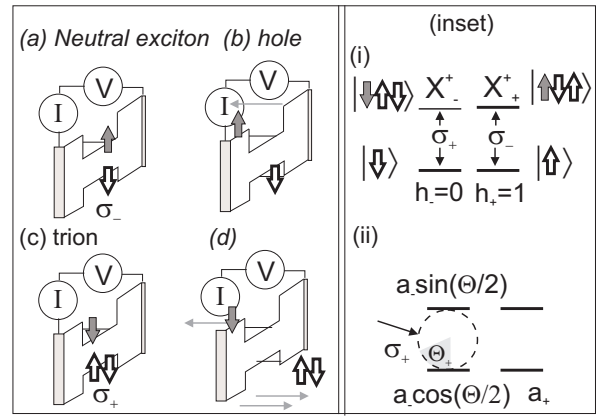}\vspace{0.2cm}
\end{center}
\caption{ Illustration of operating principle. {\it Preparation:}
~ (a) Resonant excitation of the $0-X^0$ transition creates a spin
polarized electron-hole pair. Filled (open) arrows are electron
(hole) respectively. (b)Under applied electric-field the electron
tunnels from the dot, leaving a spin-polarized hole. {\it
Read-out:} (c) A circularly polarized $\pi$-pulse creates a
charged exciton only if the hole is in the target spin state. (d)
Carriers tunnel from the dot, the creation of a charged exciton
resulting in a change in the photocurrent proportional to the
occupation of the target hole spin state. Inset: {\it Control} (i)
Energy-levels of heavy-hole/charged exciton system which acts as
two decoupled 2-level atoms: $h_{\pm}-X^+_{\pm}$. (ii) Driving a
Rabi rotation with $\sigma_+$ circular polarization addresses the
$h_--X^+_-$ transition only.
 }\label{fig:opprinc}
\end{figure}

First we shall describe the principle of operation. The qubit is
represented by the spin states of the heavy-hole
($J=\frac{3}{2}$), where logical states 0(1), are the spin up
(down) states ($m_J = \pm\frac{3}{2}$). Figure~\ref{fig:opprinc}
shows an idealized  quantum dot, embedded in an n-i-Schottky diode
structure. An electric-field is applied, such that the electron
tunnelling rate is much faster than the hole tunnelling rate. The
experiments use a sequence of two circularly polarized,
time-separated laser pulses, with a time-duration shorter than the
electron tunnelling time, labelled the `preparation' and `control'
pulses. Figure \ref{fig:opprinc} illustrates the steps (a-d)
involved in the preparation, and read-out of the
hole spin. \\
\indent {\it Preparation:}(a) The circularly polarized preparation
pulse resonantly excites the ground-state neutral exciton
transition ($0-X^0$), driving a Rabi rotation through an angle
equal to the pulse-area of $\pi$ \cite{Nielsen_book}. This creates
a spin-polarized electron-hole pair with near unit probability.
(b) Under the action of the applied electric-field the electron
will tunnel from the dot, resulting in a photocurrent proportional
to the final exciton population of up to one electron per pulse,
which for a 76-MHz repetition rate is 12.18 pA
\cite{Zrenner_nat,Stufler_prb}. Since the electron tunnelling rate
is much faster than for the hole, the electron tunnels from the
dot leaving a spin polarized hole.\\
\indent {\it Control:} (inset) To control the $h-X^+$ transitions,
the following control scheme is used. Because the spin lifetimes
are long compared with the duration of the control laser pulse,
the heavy-hole/charged-exciton 4-level system acts as two
decoupled 2-level atoms, as illustrated in fig.
\ref{fig:opprinc}(inset). The optical selection rules are a result
of Pauli-blocking, with each hole spin-state coupling to a single
auxiliary state: $\vert h_{\pm}\rangle -\vert X^+_{\pm}\rangle $.
For a laser on resonance with the $h-X^+$ transition the control
Hamiltonian is \cite{Nielsen_book}:

\[\hat{H}= \frac{\hbar}{2}\left( \begin{array}{cccc}
0 & \Omega_+  & 0 & 0 \\
\Omega_+ & 0 & 0 & 0 \\
0 & 0 & 0 & \Omega_- \\
0 & 0 & \Omega_- & 0 \end{array} \right)\]

\noindent where $\Omega_{\pm}$ are the Rabi frequencies of the
circular polarization components of the control laser, and the
basis is $\vert \psi\rangle=(\vert h_-\rangle, \vert X^+_-\rangle,
\vert h_+\rangle, \vert X^+_+\rangle)$. The control laser pulse
implements the unitary operation
$\hat{U}(\Theta_+,\Theta_-)=\exp{(\frac{i}{\hbar}\int H dt)}$:

\[\hat{U} =\left( \begin{array}{cccc}
\cos{(\frac{\Theta_+}{2})} & -i\sin{(\frac{\Theta_+}{2})}   & 0 & 0 \\
-i\sin{(\frac{\Theta_+}{2})} & \cos{(\frac{\Theta_+}{2})} & 0 & 0 \\
0 & 0 & \cos{(\frac{\Theta_-}{2})} & -i\sin{(\frac{\Theta_-}{2})} \\
0 & 0 & -i\sin{(\frac{\Theta_-}{2})} &
\cos{(\frac{\Theta_-}{2})}\end{array} \right)\]

\noindent  where $\Theta_{\pm}=\int \Omega_{\pm}dt$ are the
pulse-areas of the circularly polarized laser components. Control
of the phase of the hole spin could be achieved as follows. A
$\sigma_+$ polarized control laser addresses one transition only.
For an initial-state of $\vert\psi\rangle=(a_-,0,a_+,0)\equiv
(a_-,a_+)$, a $\sigma_z$-gate imparting a relative phase-shift of
$\pi$ between the hole spin states would be implemented when
$\Theta_-=0,~\Theta_+=2\pi$:

\[\hat{U}\rightarrow\hat{\sigma}_z=\left( \begin{array}{cc}
1 & 0\\
0 & -1 \\
\end{array} \right)\]

The phase-shift arising from a 2$\pi$ Rabi-rotation has been
verified in four-wave mixing experiments on the neutral exciton of
an interface dot \cite{Patton_prl}. Further discussion of this
control scheme can be found in refs.
\cite{Nielsen_book,Lovett_prl}.

{\it Read-out:} (c-d) Creation of the charged exciton results in a
change in the photocurrent signal, and hence a read-out
proportional to the probability that the hole is in the target
spin state at that instant in time.

Full details of the device can be found in ref.
\cite{Kolodka_prb}, where inversion recovery measurements on this
dot confirm that the neutral exciton coherence is limited by
electron tunnelling. Due to the electron-hole exchange interaction
the exciton transitions are linearly polarized and have a
fine-structure splitting of $h/(230\pm 10~\mathrm{ps})$. Resonant
excitation in step (a) with circular polarization creates a
spin-polarized exciton. A combination of the fine-structure beat,
and the time for the electron to tunnel leads to some loss of spin
polarization. However, at the reverse bias of 0.8~V used in the
experiment, the electron tunnelling rate
($\Gamma_e^{-1}=35-40~\mathrm{ps}$) is  fast compared with the
period of the fine-structure beat, minimizing any loss of spin
orientation, and is slow enough to observe weakly damped Rabi
oscillations (see later). At the same time the slow hole
tunnelling rate ($\Gamma_h^{-1}\sim \mathrm{ns}$) is much faster
than the repetition rate of the laser ensuring the dot is
initially in the crystal ground-state.

\begin{figure}
\begin{center}
\includegraphics[scale=0.8]{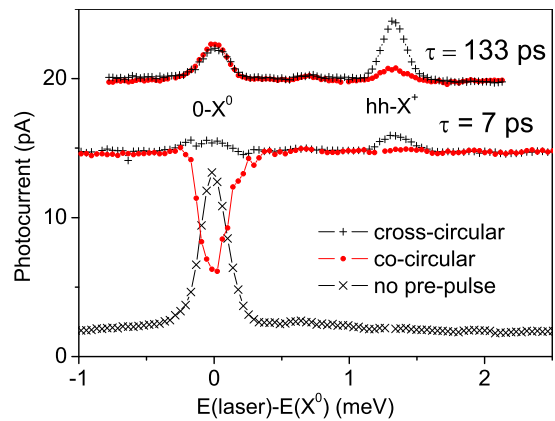}\vspace{0.2cm}
\end{center}
\caption{ Photocurrent vs laser detuning. (lower) Single circular
polarized $\pi$-pulse, with $0-X^0$ peak. (middle,upper) Two pulse
spectra, with preparation-pulse resonant with $0-X^0$ transition,
with short (long) time-delay. The upper trace is offset for
clarity. Note the emergence of the $h-X^+$ line at longer
time-delays for cross-polarized excitation. }\label{fig:detune}
\end{figure}


Figure \ref{fig:detune} presents one and two color photocurrent
spectra to show the preparation and detection of a single hole
spin. The lower trace of fig.~\ref{fig:detune} presents the case
of single pulse excitation. A single peak corresponding to the
neutral exciton transition ($0-X^0$) is observed, with lineshape
determined by the Gaussian pulse shape (FWHM=0.2 meV).

In the case of two color excitation, a preparation pulse with
pulse-area $\pi$ is tuned to the $0-X^0$ transition, to create a
neutral exciton with a probability close to one. The photocurrent
is then recorded as a function of the detuning of the
control-pulse, which also has a pulse-area of $\pi$. The middle
traces of fig.~\ref{fig:detune} show two-color photocurrent
spectra for a   time-delay of
 7~ps, much shorter than the electron tunnelling
time. For  co-polarized pulses there is a dip at the ($0-X^0$)
transition, since the pulse-pair is now equivalent to a
$2\pi$-pulse. For the cross-polarized case there is only a very
weak ($0-X^0$) feature, but importantly there is an additional
peak at a detuning of $+1.32~\mathrm{meV}$, corresponding to the
heavy-hole to charged exciton transition ($h-X^+$)
\cite{Ware_prl}. As the time-delay increases the electron tunnels
from the dot, and the heavy-hole population grows, as  seen in the
upper traces of fig.~\ref{fig:detune}. At a time-delay of 133~ps,
which is much longer than the 35-40 ps electron tunnelling time,
the exciton is completely ionized, resulting in a weak
polarization insensitive ($0-X^0$) peak, and a stronger
polarization sensitive $h-X^+$ peak.

The ($0-X^0$) and the ($h-X^+$) features in the two pulse spectra
have the opposite selection rules. For ($0-X^0$), the Coulomb
interaction shifts the energy of the biexciton by
$-2.16~\mathrm{meV}$ and out-of-resonance with the spectrally
narrow laser pulse preventing the absorption of the
cross-polarized control-pulse. In the case of the positive trion,
absorption of a co-polarized pulse is forbidden by the Pauli
exclusion principle, since it would result in two holes of the
same spin, as shown in fig.~\ref{fig:opprinc}(inset). By contrast
cross-polarized excitation of the positive trion results in a
change in photocurrent proportional to the occupation of the
target hole spin state. The energy separation between the $X^0$
and $X^+$ transitions is in close agreement with PL
measurements. 

From the amplitudes of the $h-X^+$ peaks for cross (4.2~pA) and
co-polarized (0.88~pA) excitation at a  time-delay of 133~ps, we
deduce that when there is a hole, there is at least an 83\%
probability of the hole occupying the desired spin state. At 133
ps, there is also a
 $0-X^0$ peak, indicating an approximately 20\%
probability of the dot occupying the crystal ground-state,
implying that no hole of either spin has been prepared, possibly
due to radiative recombination of the neutral exciton. This
demonstrates steps a-d in fig. \ref{fig:opprinc}, showing
preparation, and detection of a single spin.

\begin{figure}
\begin{center}
\includegraphics[scale=1.5]{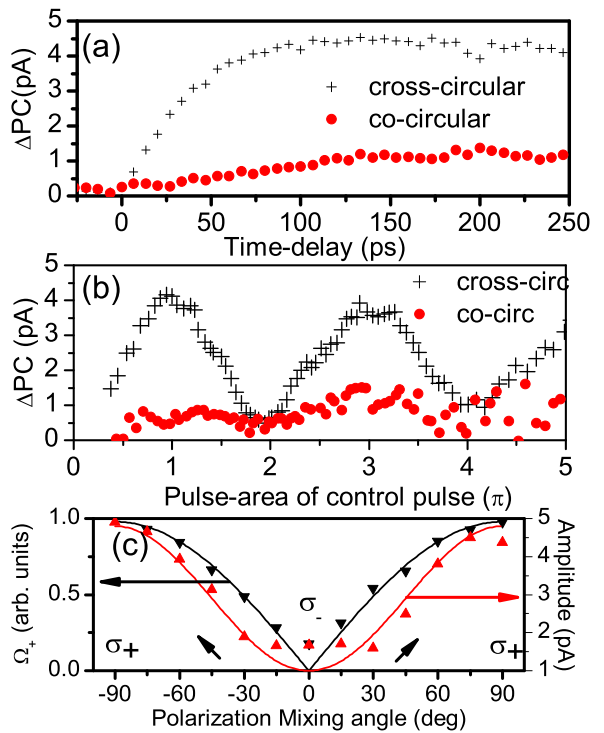}\vspace{0.2cm}
\end{center}
\caption{ (a) Time-resolved measurements of heavy-hole population.
Photocurrent vs time-delay between pre-pulse, and a control-pulse
resonant with the $h-X^+$ transition. (b)Rabi rotation of the
$h-X^+$ transition, where creation of the charged exciton is
conditional on the heavy-hole spin state. 
The $\pi$-pulse provides a read-out of the hole spin state.
(c) Polarization dependence
$(\Omega_-,\Omega_+)\equiv\Omega[\cos(\alpha),\sin(\alpha)]$ of
the $h-X^+$ Rabi rotation. ($\blacktriangledown$) Inverse period
of $h-X^+$ Rabi rotation vs polarization angle $\alpha$ of the
control pulse, for a $\sigma_-$ preparation pulse. The line shows
the $\vert \sin(\alpha)\vert$ dependence confirming the
independence of the $h_{\pm}-X^+_{\pm}$ transitions.
($\blacktriangle$) Amplitude of Rabi rotation for
$\sigma_+$-polarized control-pulse, versus the polarization angle
of the preparation pulse. }\label{fig:time}
\end{figure}

Figure ~\ref{fig:time}(a) presents  time-resolved measurements
from which the heavy-hole population can be deduced.  The
preparation-pulse creates a neutral exciton, whilst the
control-pulse of pulse-area $\pi$, resonant with $h-X^+$, probes
the population of the target hole spin state. For cross-circular
excitation an exponential rise is observed as the electron tunnels
from the dot, and the heavy-hole population increases until
saturation (as illustrated in figs.~\ref{fig:opprinc}(b)). The
hole population has an exponential rise time of 40-ps, consistent
with the electron tunnelling time \cite{Kolodka_prb}. After the
fast initial rise the hole population slowly decays with a
lifetime in excess of 600-ps. Due to the electron-hole exchange
interaction of the neutral exciton, the hole ends up with the
opposite spin about 20\% of the time, resulting in a slower rise
of the co-polarized signal \cite{Bfield}. This is the first
time-resolved measurement of a single quantum dot spin with
sub-nanosecond time resolution.


To demonstrate control of the $h-X^+$ transition, as depicted in
fig. \ref{fig:opprinc}(inset), we study the Rabi rotation of the
transition. Figure ~\ref{fig:time}(b) shows the photocurrent
versus pulse-area of the control pulse, at a time-delay of 133 ps.
Two pulses are incident on the sample: the preparation-pulse, and
a control pulse of variable pulse-area resonant with the $h-X^+$
transition. A background photocurrent linear in power has been
subtracted \cite{Zrenner_nat,Stufler_prb}.
For cross-circular excitation more than two periods of a weakly
damped  Rabi oscillation are observed. 
 For co-circularly
polarized excitation, the Rabi rotation is suppressed. The results
 in fig. \ref{fig:time}(b) demonstrate a Rabi rotation of a
charged exciton conditional on the initial spin state. Previous
reports of a Rabi oscillation of a charged exciton were for  {\it
uninitialized} spins, in both an ensemble of quantum dots
\cite{Greilich_prl}, and for an excited state charged exciton of
unknown charge \cite{Besombes_prl}.

To confirm that the 4-level $h-X^+$ system behaves as two
decoupled two-level optical transitions, which are rotated by
$\hat{U}$ when excited by the control laser, we studied the
polarization dependence of the $h-X^+$ Rabi rotation. In the first
experiment a $\sigma_-$ preparation pulse is used to create an
initial state which is predominantly $\vert h_-\rangle$:
$\vert\psi\rangle\approx (1,0,0,0)$. The Rabi rotation is then
measured as a function of the polarization of the control pulse,
defined as:
 $(\Omega_-,\Omega_+)\equiv\Omega[\cos(\alpha),\sin(\alpha)]$.
 The amplitude of the Rabi rotation
 is almost constant, but the
inverse period is equal to the $\vert\sin{(\alpha)}\vert$
amplitude of the $\sigma_+$ component of the Rabi frequency of the
control pulse, as seen in fig. \ref{fig:time}(c). This
demonstrates that the $\vert h_{\mp}\rangle$ state only interacts
with $\sigma_{\pm}$ polarized light.

In the second experiment, the polarization of the control pulse is
fixed at $\sigma_-$, and the Rabi rotation is measured as a
function of the polarization of the preparation pulse. The period
of the Rabi rotation is constant, but the amplitude exhibits a
$\sin^2{(\alpha)}$ dependence reflecting the occupation of the
$\vert h_-\rangle$ state, as shown in fig. \ref{fig:time}(c). This
demonstrates that the polarization of the preparation pulse can be
used to control the initial populations of the hole spin states
$\vert h_{\pm}\rangle$, in the mixed state. The distinct
$\vert\sin{(\alpha)}\vert$ and $\sin^2{(\alpha)}$ dependencies of
these measurements, further confirm that $\hat{U}$ is a good
approximation of the action of the control pulse.

Armed with tools for the sequential initialization and read-out of
a single spin, a number of future experiments are now possible.
For example, a magnetic field in the Voigt configuration may be
used to achieve an arbitrary phase-shift on a single spin
\cite{spin_gates}. To observe the precession of the hole spin a
preparation and read-out pulse sequence would be applied. The data
presented here strongly suggests that when a third circularly
polarized control pulse with a pulse-area of $2\pi$ is applied,
an operation $\hat{U}(0,2\pi)\approx \hat{\sigma}_z$ will induce a
relative phase-shift of $\pi$ between the hole spin states,
resulting in a phase-jump in the spin precession. The phase-shift
can then be controlled using the detuning of the laser
\cite{Gauger_arxiv}.

To summarize, using a photodiode structure we demonstrate
sequential initialization, coherent optical control, and
photocurrent read-out of a single hole spin.  This work
establishes an experimental platform for investigating optical
control of single quantum dot spins, which marries the ultrafast
coherent control of excitons with the long coherence times of spin
based qubits.

This work was funded by EPSRC UK GR/S76076 and the QIPIRC UK.
J.~B.~B.~O acknowledges financial support from CAPES Brazil.
Following submission of this work, Mikkelsen {\it et al}
\cite{Mikkelsen_natphys} reported the coherent precession of a
single electron spin.

\bibliographystyle{apsrev}
\bibliography{Ramsay_080309}

\begin{thebibliography}{1}

 

\bibitem{Loss_pra}
D.~Loss and D.~P.~DiVincenzo, \pra {\bf 57} 120 (1998).

\bibitem{Imamoglu_prl}
A.~Imamoglu, D.~D.~Awschalom, G.~Burkard, D.~P.~DiVincenzo, D.~ Loss, M.~Sherwin, 
and A.~Small, \prl {\bf 83} 4204 (1999).

\bibitem{Zrenner_nat}
A.~Zrenner, E.~Beham, S.~Stufler, F.~Findeis, M. Bichler, and G. Abstreiter,
 Nature {\bf 418}, 612 (2002).

\bibitem{Stufler_prb}
S.~Stufler, P.~Ester, A.~Zrenner, and M.~Bichler, \prb {\bf 72} 121301(R) (2005).

\bibitem{Li_sci}
X.~Li, Y.~Wu, D.~G.~Steel, D.~Gammon, T.~H.~Stievater, D.~S.~Katzer, D.~Park, C.~Piermarocchi, and L.~J.~Sham
Science {\bf 301}, 809 (2003).

\bibitem{Patton_prl}
B.~Patton, U.~Woggon, and W.~Langbein, \prl {\bf 95} 266401 (2005).   


\bibitem{Kroutvar_nat}
M.~Kroutvar, Y.~Ducommun, D.~Heiss, M.~Bichler, D.~Schuh, G.~Abstreiter, J.~J.~Finley, Nature {\bf 432} 81 (2004).

\bibitem{Heiss_cmat}
D.~Heiss, S. Schaeck, H.~Huebl, M.~Bichler, G.~Abstreiter, J.~J.~Finley, D.~V.~Bulaev, and D.~Loss,
\prb {\bf 76} 241306(R) (2007).

\bibitem{Greilich_sci}
A.~Greilich, D.~R.~Yakovlev, A.~Schabaev, AI.~L.~Efros, I.~A.~Yugova, R.~Oulton, V.~Stavarache, 
D.~Reuter, A.~Wieck, and M.~Bayer, Science, {\bf 313}, 341 (2006). 







\bibitem{Atature_sci}
M.~Atature, J.~ Dreiser, A.~ Badolato, A.~ Hogele, K.~ Karrai, and A.~ Imamoglu, Science {\bf 312} 515 (2006).  


\bibitem{Xu_prl}
X.~Xu, Y.~Wu, B.~Sun, Q.~Huang, J.~Cheng, D.~G.~Steel, A.~S.~Bracker, D.~Gammon, 
C.~Emary, and L.~J.~Sham, \prl {\bf 99}, 097401 (2007).

\bibitem{Gerardot_nat}
B.~D.~Gerardot, D.~Brunner, P.~A.~Dalgarno, P.~\"{O}hberg, S.~Seidl, M.~Kroner, K.~Karrai, 
N.~G.~Stoltz, P.~M.~Petroff, and R.~J.~Warburton, Nature {\bf 451} 441 (2008).

\bibitem{Berezovsky_sci}
J.~Berezovsky, M.~H.~Mikkelsen, O.~Gywat, N.~G.~Stoltz, L.~A.~Coldren, and D.~D.~Awschalom, Science,
{\bf 314} 1916 (2006).

\bibitem{Atature_natphys}
M.~Atature, J.~ Dreiser, A.~ Badolato, and A.~ Imamoglu, Nat. Phys. {\bf 3} 101 (2007).  

\bibitem{Greilich_prl}
A.~Greilich, R.~Oulton, E.~A.~Zhukov, I.~A.~Yugova, D.~R.~Yakovlev, M.~Bayer, A.~Shabaev, AI.~L.~Efros,  
 I.~A.~Merkulov, V.~Stavarache, 
D.~Reuter, and A.~Wieck,  \prl, {\bf 96}, 227401 (2006). 

\bibitem{Wu_prl}
Y.~Wu, E.~D.~Kim, X.~Xu, J.~Cheng, D.~G.~Steel, A.~S.~Bracker,
D.~Gammon, S.~E.~Economou, and L.~J.~Sham, \prl {\bf 99} 097402 (2007). 

\bibitem{Nielsen_book}
M.~A.~Nielsen and I.~L.~Chuang {\it Quantum computation and Quantum information} 
(Cambridge University Press, Cambridge, 2000), p279, p303 and p319.

\bibitem{Lovett_prl}
A.~Nazir, B.~W.~Lovett, S.~D.~Barrett, T.~P.~Spiller, and G.~A.~D.~Briggs, 
\prl {\bf 93} 150502 (2004).


\bibitem{Kolodka_prb}
R.~S.~Kolodka, A.~J.~Ramsay, J.~Skiba-Szymanska, P.~W.~Fry, H.~Y.~Liu,
 A.~M.~Fox, and M.~S.~Skolnick, \prb {\bf 75} 193306 (2007).



\bibitem{Ware_prl}
M.~E.~Ware, E.~A.~Stinaff, D.~Gammon, M.~F.~Doty, A.~S.~Bracker,
D.~Gershoni, V.~L.~Korenev, \c{S}.~C.~B\u{a}descu, Y.~Lyanda-Geller,
and T.~L.~Reinecke,  \prl {\bf 95} 177403 (2005).

\bibitem{Bfield}
Spin scattering due to electron-hole exchange could be reduced by applying a B-field
parallel to the growth direction.



\bibitem{Besombes_prl}
L.~Besombes, J.~J.~Baumberg and J.~Motohisa, \prl {\bf 90}, 257402 (2003).





























































\bibitem{spin_gates}
S.~E.~Economou, and T.~L.~Reinecke, \prl, {\bf 99} 217401 (2007).

\bibitem{Gauger_arxiv}
E.~M.~Gauger, S.~C.~Benjamin, and B.~W.~Lovett, Quant ph:- 0708.1692 (2007).



\bibitem{Mikkelsen_natphys}
M.~H.~Mikkelsen, J.~ Berezovsky, N.~G.~Stolz, L.~A.~Coldren, and D.~D.~Awschalom, 
Nature Physics, {\bf 3}, 771 (2007).


















\end{thebibliography}

\end{document}